\documentclass[letterpaper]{article} 
\usepackage{aaai2026}  
\usepackage{times}  
\usepackage{helvet}  
\usepackage{courier}  
\usepackage[hyphens]{url}  
\usepackage{graphicx} 
\usepackage{natbib}  
\usepackage{caption} 
\usepackage{algorithm}
\usepackage{algorithmic}
\usepackage{amsmath}
\usepackage{amssymb}
\usepackage{makecell}
\usepackage{booktabs}
\usepackage{multirow}
\usepackage{pifont}
\usepackage{xcolor}
\newcommand{\cmark}{{\ding{52}}} 
\newcommand{\xmark}{{\ding{56}}} 

\usepackage{newfloat}
\usepackage{listings}
\DeclareCaptionStyle{ruled}{labelfont=normalfont,labelsep=colon,strut=off} 
\floatstyle{ruled}
\newfloat{listing}{tb}{lst}{}
\floatname{listing}{Listing}
\title{MuSpike: A Benchmark and Evaluation Framework for Symbolic Music Generation with Spiking Neural Networks}
\author{
    Qian Liang\textsuperscript{\rm 1}\\
    Menghaoran Tang\textsuperscript{\rm 2}\equalcontrib,
    Yi Zeng\textsuperscript{\rm 1,2,3}\thanks{Corresponding Author.}
}
\affiliations{
    \textsuperscript{\rm 1}Brain-inspired Cognitive Intelligence Lab, Institute of Automation, Chinese Academy of Sciences\\
    \textsuperscript{\rm 2}University of Chinese Academy of Sciences, Beijing, China \\
    \textsuperscript{\rm 3} Key Laboratory of Brain Cognition and Brain-inspired Intelligence Technology, Chinese Academy of Sciences, China.

    qian.liang@ia.ac.cn, tangmenghaoran22@mails.ucas.ac.cn, yi.zeng@ia.ac.cn
}

\usepackage{bibentry}

\begin{document}

\maketitle

\begin{abstract}
Symbolic music generation has seen rapid progress with artificial neural networks, yet remains underexplored in the biologically plausible domain of spiking neural networks (SNNs), where both standardized benchmarks and comprehensive evaluation methods are lacking. To address this gap, we introduce MuSpike, a unified benchmark and evaluation framework that systematically assesses five representative SNN architectures (SNN-CNN, SNN-RNN, SNN-LSTM, SNN-GAN and SNN-Transformer) across five typical datasets, covering tonal, structural, emotional, and stylistic variations. MuSpike emphasizes comprehensive evaluation, combining established objective metrics with a large-scale listening study. We propose new subjective metrics, targeting musical impression, autobiographical association, and personal preference, that capture perceptual dimensions often overlooked in prior work. Results reveal that (1) different SNN models exhibit distinct strengths across evaluation dimensions; (2) participants with different musical backgrounds exhibit diverse perceptual patterns, with experts showing greater tolerance toward AI-composed music; and (3) a noticeable misalignment exists between objective and subjective evaluations, highlighting the limitations of purely statistical metrics and underscoring the value of human perceptual judgment in assessing musical quality. MuSpike provides the first systematic benchmark and systemic evaluation framework for SNN models in symbolic music generation, establishing a solid foundation for future research into biologically plausible and cognitively grounded music generation.
\end{abstract}

\begin{links}
    \link{Code}{https://github.com/lqnankai/MuSpike.git}
\end{links}

\section{Introduction}

Music is regarded as the intersection of structure and emotion, combining time-based patterns with expressive meanings that go beyond language. Symbolic music generation, which represents music using discrete symbols, has emerged as a key direction in AI research, focusing on high-level musical features such as melodic contour, harmonic progression, rhythmic structure, and stylistic variation. These features make it especially suitable for learning musical structure and semantics through neural sequence models. In recent years, significant progress has been made through deep learning models such as recurrent neural networks (RNNs) and long short-term memory networks(LSTMs)~\cite{briot2020deep, eck2002first, magenta2016, boulanger2012modeling,hadjeres2017deepbach}, generative adversarial networks (GANs)~\cite{JazzGAN,dong2018musegan},  Transformers~\cite{huang2020pop, MuseNet, Zhao2024AdversarialMidiBERTSM, xu2024generating}, even pre-trained large language model~\cite{wang2025notagen}. These models are capable of learning rich temporal dependencies in music and have been successfully applied to diverse generation tasks.

However, music is inherently temporal, structured by evolving hierarchies of melody, rhythm, and harmony. As a product of human cognition, it demands models capable of processing temporal information precisely and encoding internal structure in biologically meaningful ways. \textbf{Spiking Neural Networks (SNNs)}, which simulate the dynamics of biological neurons through time-dependent spike-based communication, inherently support asynchronous, event-driven computation with high temporal precision. These properties make SNNs a promising yet underexplored candidate for symbolic music generation. To date, this area lacks a standardized benchmark for comparing different SNN architectures, leaving their generative capabilities insufficiently understood. 

Moreover, evaluating generated music generally follows two paradigms: objective statistical metrics and human subjective assessments. While both approaches are widely adopted, existing studies in both the ANN and SNN domains tend to apply them in isolation or without consistency. Currently, there is no comprehensive framework that integrates objective and subjective evaluations, nor is there sufficient exploration of the relationship between them. This challenge not only undermines fair and reproducible comparison across models, but also limits our understanding of how AI-generated music aligns with human cognitive perception.

To address these challenges, we present \textbf{MuSpike}, the first benchmark and comprehensive evaluation framework specifically designed for symbolic music generation using spiking neural networks, to establish a unified foundation for investigating the generative capacity of SNN-based architectures. The contributions of our study are as follows:
\begin{itemize}
    \item We propose \textbf{MuSpike}, a standardized benchmark that supports five representative SNN architectures, Spiking CNN (S-CNN), Spiking RNN (S-RNN), Spiking LSTM (S-LSTM), Spiking GAN (S-GAN), and Spiking Transformer (S-Transformer) across five widely-used symbolic music datasets, covering tonal, structural, emotional, and stylistic variations.
    \item We introduce a comprehensive evaluation framework that integrates objective statistical metrics with cognition-informed subjective assessments, and propose novel cognition-level subjective metrics, including musical impression, autobiographical association, and personal preference, that capture perceptual and psychological aspects of musical experience often overlooked in prior work.
    \item For convenience and reproducibility, we develop a large-scale listening test platform that enables multidimensional evaluation of generated music. Based on this platform, we systematically analyze the relationship between objective and subjective evaluation paradigms, revealing both alignment and divergence between statistical regularities and human perceptual judgments.
\end{itemize}

\section{Related Works}
\subsection{Spiking Neural Networks in Music Tasks}
Spiking Neural Networks have gained increasing attention in recent years. However, symbolic music generation with SNNs remains significantly underexplored. Recently, a few researchers have begun to investigate the potential of SNNs in generative music tasks. A brain-inspired sequential memory model has been developed to encode, store, and retrieve musical fragments~\cite{LQ2020}. Building on this memory foundation, subsequent studies have explored stylistic composition~\cite{LQ2021}, emotional generation~\cite{BrainCog}, and mode- and key-conditioned learning and four-part music generation~\cite{LQ2025}, demonstrating the feasibility of biologically inspired temporal modeling in symbolic music tasks. In parallel, SpikingMuseGAN was proposed to leverage spiking neurons for generating high-quality emotional music in a controllable manner~\cite{spikemusegan}. Another work explored melody generation by replacing the standard LSTM units with two biologically inspired variants of leaky integrate-and-fire (LIF) neurons, showing the potential of SNN-based recurrent structures in music sequence modeling.~\cite{10.1117/12.2659783}.
\subsection{Diversity of Symbolic Music Datasets }
A variety of symbolic music datasets have been developed to support different sequence modeling and generative tasks in the symbolic music domain. 
For instance, the \textbf{J. S. Bach’s four-part chorales (JSB)} dataset is a longstanding benchmark for polyphonic modeling and harmonic analysis~\cite{cuthbert2010music21}. The \textbf{Lakh MIDI Dataset}~\cite{bertin2011million} spans multiple genres, including pop, jazz, and classical, and is aligned with the Million Song Dataset. \textbf{POP909}~\cite{wang2020pop909} contains annotated pop songs with melody, chord, and phrase-level labels, making it suitable for structure-conditioned music generation. \textbf{EMOPIA}~\cite{hung2021emopia}, annotated with four-quadrant emotion labels based on the Russell’s valence-arousal model of affect~\cite{russell1980circumplex}, has been widely used in affective music generation tasks. \textbf{XMIDI}~\cite{tian2025xmusic} offers consistently formatted MIDI files with emotion and genre labels, along with expressive annotations. The \textbf{Nottingham Music Dataset (NMD)}~\cite{Nottingham} is a collection of British and American folk melodies with corresponding chord annotations. \textbf{Piano-MIDI}, \textbf{SHTE}~\cite{LQ2025} and other larger corpora such as \textbf{GiantMIDI}~\cite{Kong_2022}, \textbf{Classical Archives}~\cite{pianomidi}, and the \textbf{TheoryTab Dataset}~\cite{theorytab} extend coverage to Western classical and popular music, supporting analysis of tonal structure, harmonic progression, and form for different generation purposes.

\subsection{Challenges in Evaluation}
Evaluating the quality of symbolic music generation remains a longstanding and multifaceted challenge. Existing works typically rely on objective and subjective evaluation methods~\cite{xiong2023}. 

\subsubsection{Objective evaluation} offers quantitative insights into the structural properties of generated music by analyzing its statistical and structural properties. Ji et al.~\cite{ji2023survey} conducted a comprehensive survey and summarized previous studies, categorizing these metrics into three groups: pitch-related, rhythm-related and harmony-related. To facilitate standardized evaluation, several toolkits have been proposed. Yang et al.~\cite{yang2020evaluation} released an open-sourced toolbox that computes absolute and relative pitch- and rhythm-based metrics using measures such as Kullback–Leibler divergence and overlapped area. Muspy~\cite{dong2020muspy}, a widely used symbolic music processing library, provides a variety of objective metrics such as polyphony, pitch entropy, and rhythm-based features like empty-beat ratio and groove consistency. These tools have significantly lowered the barrier for conducting reproducible evaluation.
\subsubsection{Subjective evaluation} is widely regarded as the most direct and ecologically valid approach for assessing the quality of symbolic music generation~\cite{cowen2020music}. Music has the capacity to evoke complex emotional states~\cite{juslin2008emotional,koelsch2014brain}, trigger autobiographical memories~\cite{janata2007characterisation}, and engage deeply rooted cognitive and cultural expectations\cite{frith1996music}, which are often beyond the reach of objective computational metrics. One of the most commonly adopted approaches is the listening test, in which human participants are asked to evaluate generated musical excerpts based on predefined criteria. However, existing studies on subjective evaluation face several limitations. Some 
omit subjective evaluation altogether~\cite{spikemusegan, LQ2025}, while others rely on overly simplified protocols, such as A/B tests, Turing-style comparisons, or questionnaires with only a few subjective items, often involving small participant groups~\cite{tian2025xmusic, wang2025notagen, LQ2021}. A few studies have proposed more comprehensive evaluation schemes~\cite{hernandez2022subjective,chu2022empirical}, where participants are divided into multiple groups and asked to rate musical samples along several perceptual dimensions using Likert-scale questionnaires. Nevertheless, even these efforts often overlook deeper cognitive dimensions of music perception, such as schematic expectations, autobiographical associations, personal preferences, etc.

\begin{table}[t]
\centering
\small
\begin{tabular}{c|p{5cm}|c|c|c}
\toprule
\textbf{ID}&\textbf{Evaluation Item} & \textbf{N} & \textbf{A} & \textbf{E} \\
\midrule
Q1 & The music sounds pleasant. & \cmark & \cmark & \cmark\\
Q2 & The music sounds natural and fluent. & \cmark & \cmark & \cmark \\
Q3 & The music conveys some emotion. & \cmark & \cmark & \cmark \\
Q4 & The rhythm is consistent. & \cmark & \cmark & \cmark \\
Q5 & The music has a clear structure or repeated segments. & \xmark & \cmark & \cmark \\
Q6 & The music shows a recognizable style. & \xmark & \cmark & \cmark \\
Q7 & The music exhibits tonal coherence. & \xmark & \xmark & \cmark \\
Q8 & The harmonic progression is natural. & \xmark & \xmark & \cmark \\
Q9 & The melody exhibits melodic motivation. & \xmark & \xmark & \cmark \\
Q10 & The music sounds novel or original. & \cmark & \cmark & \cmark \\
\textbf{Q11} & \textbf{The music left a strong impression.} & \cmark & \cmark & \cmark \\
\textbf{Q12} & \textbf{The music reminded me of personal experiences}. & \cmark & \cmark & \cmark \\
\textbf{Q13} & \textbf{I like the music.} & \cmark & \cmark & \cmark \\
Q14 & Who composed it (Human / AI / Uncertain) & \cmark & \cmark & \cmark \\
\bottomrule
\end{tabular}
\caption{Subjective evaluation metrics designed for participant in Normal (N), Amateur (A) and Expert (E) groups.}
\label{tab:subjective_metrics}
\end{table}
\section{Method}
\subsection{Neuron model}
In this study, we choose the Leaky Integrate-and-Fire (LIF) neural model, one of the most widely used spiking neuron models due to its simplicity and computational efficiency. The model can be described as the equation~\ref{eq:1}:
\begin{equation}\label{eq:1}
\begin{gathered}
\tau_m \frac{dV(t)}{dt} = -V(t) + R I(t) \\
V(t) = 
\begin{cases}
    V_{\text{reset}}, & \text{if } V(t) \geq V_{\text{th}} \\
    V(t), & \text{otherwise}
\end{cases}
\end{gathered}
\end{equation}

where $V(t)$ denotes the membrane potential, $\tau_m$ is the membrane time constant, $R$ is the membrane resistance, and $I(t)$ represents the input current at time $t$. When $V(t)$ reaches a predefined threshold $V_{\text{th}}$, the neuron emits a spike and the membrane potential is reset to $V_{\text{reset}}$.
\subsection{Music Feature Encoding}
We follow the data processing pipeline proposed by Compound Word Transformer~\cite{Hsiao_Liu_Yeh_Yang_2021}, where the original MIDI sequences are converted into compound word tokens. Each token consists of seven features: six musical attributes—\textit{tempo, chord, bar-beat, position, pitch, duration}, and \textit{velocity}—plus a \textit{type} feature  that indicates the token category. The seven features are embedded, concatenated, and passed through a spike-based encoder with a linear projection and LIF neurons. These spiking representations are used as input to downstream modules. Notably, the LIF neurons are parameterized with time constant $\tau_m=2.0$ and 
firing threshold $V_{\text{th}}=0.5$. To enable gradient-based optimization, the spike encoder employs the ATan surrogate function, allowing projection weights to adaptively map symbolic features to spike trains.

\subsection{Spiking Neural Network Architectures}
We implemented five representative spiking architectures, each adapted from a well-established deep learning model that has been widely applied in symbolic music generation with artificial neural networks (ANNs): \textbf{Spiking CNN (S-CNN), Spiking RNN (S-RNN), Spiking LSTM (S-LSTM), Spiking GAN (S-GAN), and Spiking Transformer (S-Transformer)}.  The parameters and training details of each model are summarized
in Table A2 in the Appendix.

\subsection{Datasets}
To enable comprehensive and representative evaluation of symbolic music generation with spiking neural networks, we adopt five widely used datasets: \textbf{JSB Chorales}, \textbf{POP909}, \textbf{Lakh MIDI}, \textbf{EMOPIA}, and \textbf{XMIDI}. As summarized in Table~A1 (Appendix), these datasets form a balanced and representative benchmark covering tonal polyphony, phrase-level pop structure, emotional music, and large-scale genre variation.

\subsection{Evaluation Metrics}
In this section, we introduce the evaluation framework used in our benchmark, which includes both objective and subjective components designed to comprehensively assess the quality of symbolic music generation.
\subsubsection{Objective Metrics}
The objective metrics in this paper were chosen based on previous studies to ensure a balanced and widely accepted assessment of model performance, which contains the pitch-related, rhythm-related and harmony-related categories: Pitch-related metrics includes pitch count (PC), Pitch Range (PR), Average Pitch Interval(PI), Pitch Entropy (PE), Pitch Class Entropy (PCE), Pitch-in-scale rate (PSR) and Polyphony (Pol); Rhythm-related metrics contains Average inter-onset interval (IOI), Note Length Transition Matrix (NLTM), Empty-Beat Rate (EBR), Groove Consistency (GC); Harmony-related metrics comprises Pitch consonance score (PCS) and Chord tone to non-chord tone ratio (CTnCTR). The detailed description are summarized in Table A3 in the Appendix.
\begin{figure*}[t]
\centering
\includegraphics[width=0.95\textwidth]{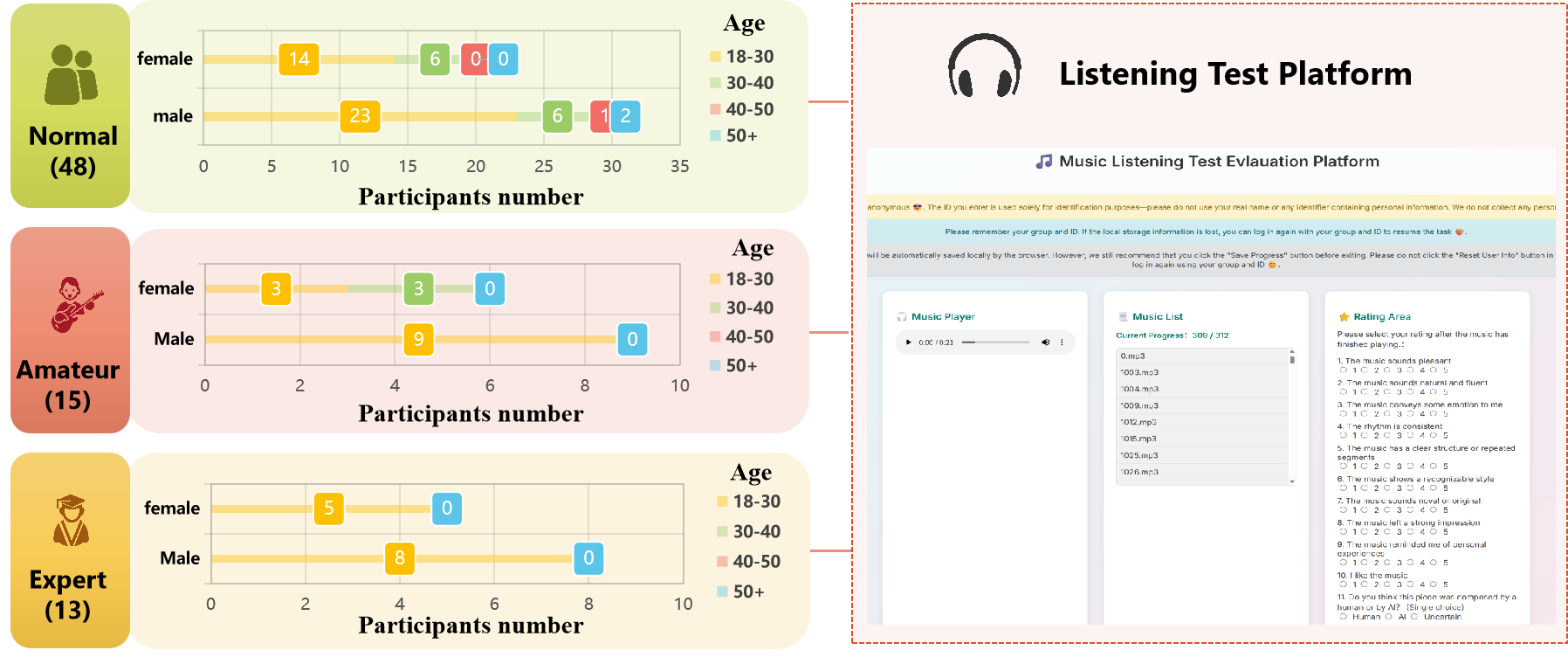} 
\caption{Participant demographics across three listener groups (Normal, Amateur, Expert) and interface of the online music listening test platform used for collecting subjective evaluations.}
\label{Fig:userstudy}
\end{figure*}
\subsubsection{Subjective Metrics}
Table~\ref{tab:subjective_metrics} summarizes the subjective evaluation metrics used in this study, covering functional dimensions such as musical fluency and coherence, emotional expressiveness, and structural attributes including tonality, harmonic progression, and thematic development. Notably, compared to most prior studies, we additionally introduce three cognitive-level metrics, impression, autobiographical association, and personal preference (corresponding to Q11, Q12, and Q13), to capture deeper cognitive dimensions of musical perception beyond those considered in prior work.

\section{User Study}\label{user}
We conducted a comprehensive user study to assess perceptual and cognitive aspects of music quality. The subjective criteria defined in Table~\ref{tab:subjective_metrics} are employed to enable cross-model comparison from a human-centered perspective.

\subsection{Participants}
A total of 96 volunteers were initially recruited for the listening test. After data validation to remove incomplete or inconsistent responses, 76 valid participants (31 female, 45 male) remained for final analysis. Participants were categorized into three groups based on their musical experience: \textbf{Normal} listeners (48 subjects) had no formal training; \textbf{Amateur} musicians (15 subjects) had learned to play at least one instrument and possibly held amateur certifications; and \textbf{Expert} musicians (13 subjects) had formal academic training in music, including students and professionals with degrees in music composition or related disciplines. Participants were also analyzed by gender and age. To ensure accessibility and reproducibility, we developed a secure online subjective evaluation platform that enabled participants to complete the listening test remotely. It is important to note that all evaluations were conducted under blind listening conditions, and participants were unaware of whether each piece was human- or AI-generated. Each participant was asked to rate 9, 11, or 14 questions based on a Likert scale, depending on their group. All participants were compensated with \$14 for their time. Summary statistics of the user study are presented in Figure~\ref{Fig:userstudy}.

\subsection{Study Design}
To construct the evaluation set for the user study, we curated a listening dataset consisting of 810 musical pieces, including 750 AI-generated and 60 human-composed samples. Specifically, for each of the five datasets, we selected 30 generated pieces from each of the five models (5 models × 5 datasets × 30 pieces = 750), and additionally sampled 12 human-composed pieces per dataset (5 datasets × 12 = 60), all drawn from the respective training sets. 

To reduce participants' listening fatigue while balancing data validity and evaluation robustness, we assigned each participant a subset of the full evaluation set under a controlled sampling strategy based on our open subjective evaluation platform. This approach ensures that each piece was evaluated a fixed number of times across participant groups. Specifically, each musical piece was evaluated at least 24 times in total, including a minimum of 16 evaluations from participants in the normal group and 4 from each of the intermediate and expert groups.  In addition, the platform supports session persistence, which enables participants to pause their evaluation and resume seamlessly, thereby alleviating listener fatigue. 
\begin{figure*}[t]
\centering
\includegraphics[width=1.0\textwidth]{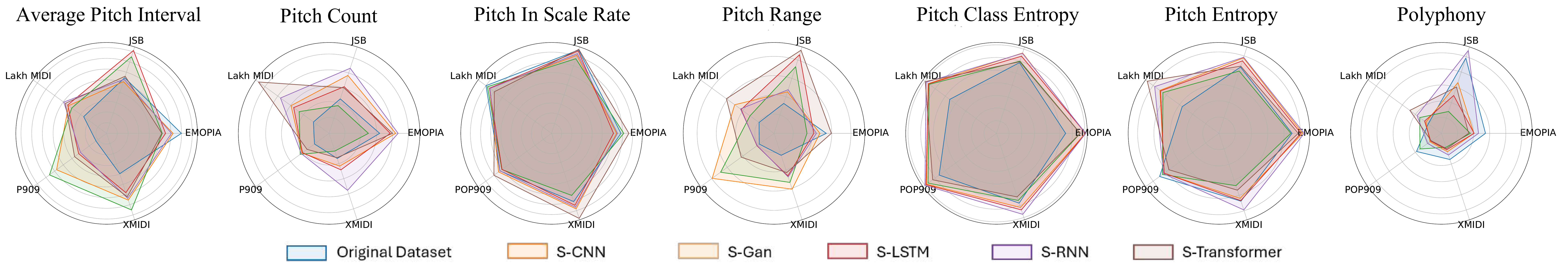} 
\caption{The results of pitch-related metrics for different Spiking models based on five datasets utilized in this paper.}
\label{Fig:obj_radar}
\end{figure*}
To further alleviate participant workload, all musical excerpts were uniformly trimmed to a maximum duration of 30 seconds. This design reduced the total listening and rating time per participant to approximately 2.5 hours. To ensure flexibility and accommodate individual schedules, each participant was allotted six days to complete the evaluation. Overall, our design effectively balanced fatigue mitigation with sample coverage and statistical validity.

\section{Experiments and Discussion}
In this section, we evaluate both the objective and subjective performance of the SNN models introduced in this paper, and further discuss the relationship between these two evaluation paradigms.

\subsection{Objective Evaluation Results}
We randomly selected 50 samples for each model on each dataset, and we computed the mean and standard deviation for all metrics. These values were also calculated for the corresponding original datasets for comparison. 

Figure~\ref{Fig:obj_radar} presents the mean values across all pitch-related metrics. The details are listed in Table A4 in the Appendix. The results show that S-RNN consistently exhibits the highest pitch diversity across datasets, with pitch count reaching $43.1$ on JSB and $37.9$ on Lakh MIDI, significantly exceeding the original datasets. This suggests that S-RNN tends to produce pitch overflow. In contrast, S-GAN shows limited pitch variety (only $18.4$ on JSB and $23.3$ on Lakh MIDI), indicating simpler or more repetitive pitch structures. Meanwhile, S-CNN and S-RNN also get high values in Pitch Entropy, reflecting higher uncertainty and variability in pitch usage. However, none of the models fully matched the high polyphony and low pitch entropy found in the original data, suggesting a gap in structural complexity and tonal coherence.
\begin{figure}[t]
\centering
\includegraphics[width=1.0\columnwidth]{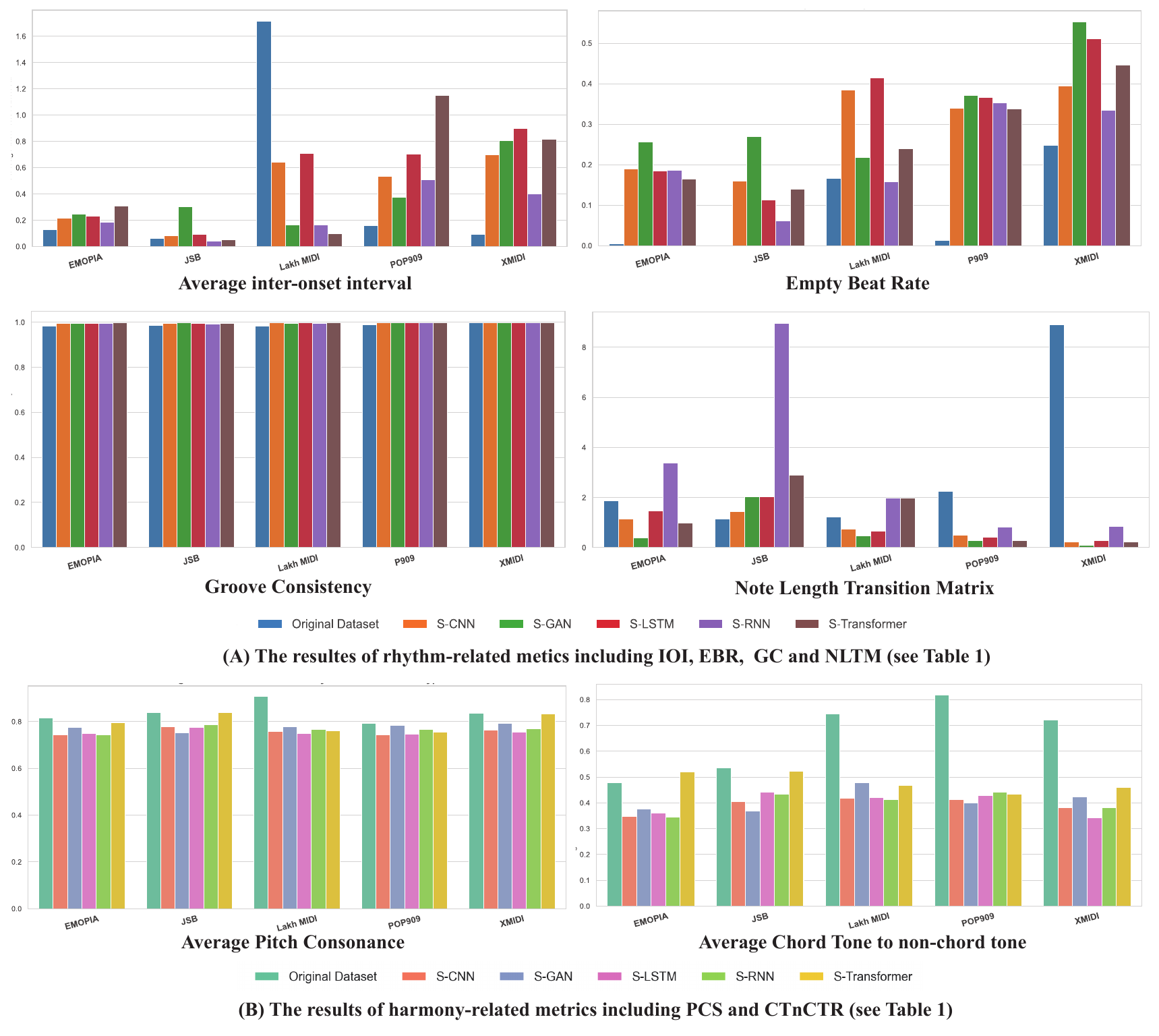} 
\caption{overview of rhythm-related and harmony-related evaluation of different Spiking models utilized based on five datasets.}
\label{Fig:rhythm-chord}
\end{figure}
Figure~\ref{Fig:rhythm-chord} and Table A5 in the Appendix describe the rhythm-related metrics between model-generated and original data. In the EMOPIA dataset, the Empty Beat Rate for original music is extremely low ($0.004$), while models like S-GAN reach as high as $0.256$, indicating excessive rhythmic gaps. Similarly, Note Length Transition in XMIDI dataset significantly decreased from $8.909$ (original) to as low as $0.102$ for S-GAN, suggesting poor rhythmic variability. Furthermore, models often overestimate IOI, such as S-Transformer in POP909, reaching $1.474$ compared to the original $0.157$. Interestingly, all models achieve groove consistency scores similar to those of the original datasets, showing an opposite pattern compared to other evaluation metrics.

For harmony-related evaluation, as shown in Figure~\ref{Fig:rhythm-chord} and Table A6, across all datasets, model-generated pieces consistently underperform human-composed references in both chord structure coherence and melodic consonance. While S-Transformer achieves relatively high PCS ($0.839$ on JSB vs. $0.837$ in original datasets), indicating its ability to produce harmonically pleasant intervals, it exhibits lower CTnTR. Other models, such as S-RNN and S-CNN, show similar patterns, with modest PCS values and markedly reduced CTnCTR scores. These results highlight that, although current models can capture harmonic consonance to some extent, they struggle to maintain structural harmonic consistency, underscoring a key limitation in learning and generating coherent harmonic progressions.

\subsection{Subjective Evaluation Results}
In this section, we will analyze the details of the results of the subjective evaluation from three groups of participants along multiple dimensions.
\begin{figure*}
\centering
\includegraphics[width=1.0\textwidth]{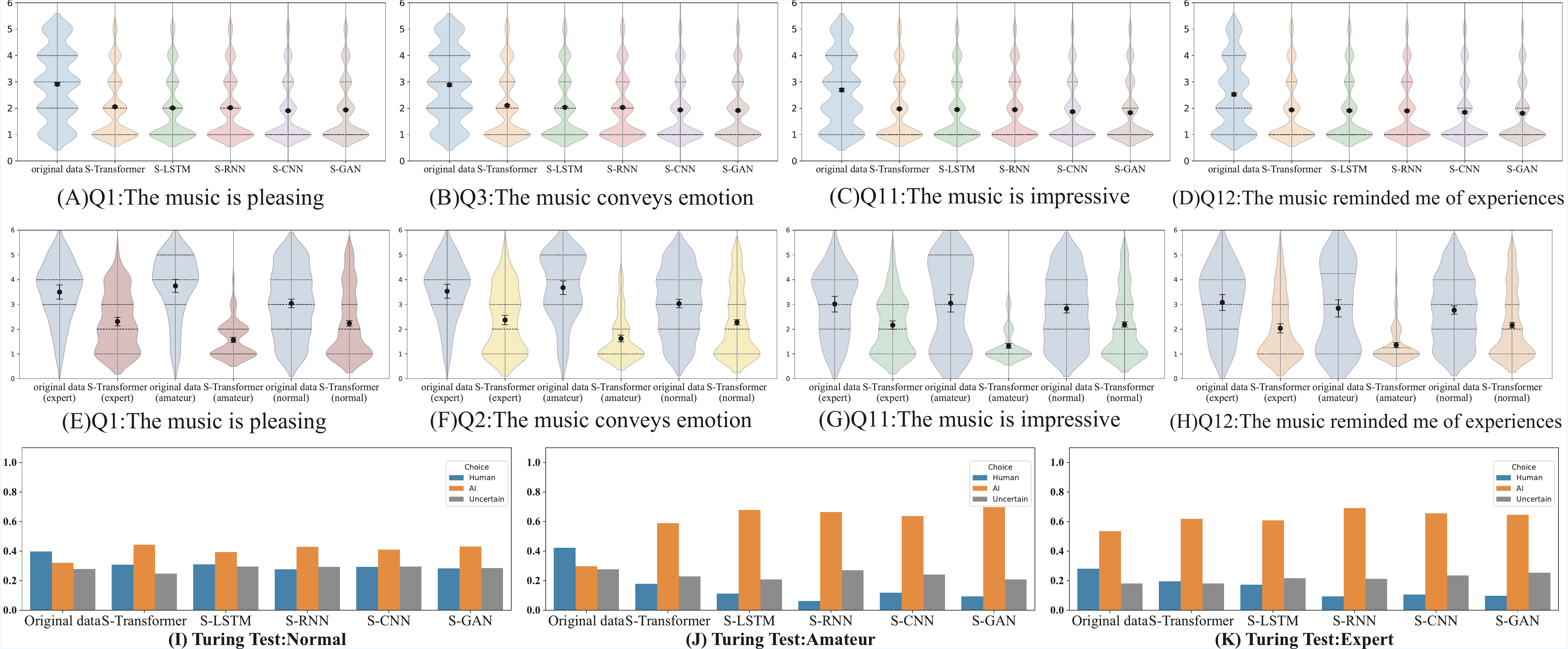} 
\caption{Overview of subjective evaluation results. (A)–(D) present the overall performance of different spiking models on key perceptual metrics: Q1, Q3, Q11, and Q12 (see Table~\ref{tab:subjective_metrics}); (E)–(H) illustrate inter-group variations across these metrics; (I)–(K) show the Turing test results for participants in each listener group.}
\label{Fig:total-group}
\end{figure*}
\begin{table*}[t]
\centering
\begin{tabular}{p{6.2cm}cccccc}
\toprule
Question & \textit{Reference} & S-Transformer & S-LSTM & S-RNN & S-GAN & S-CNN \\
\midrule
The music sounds pleasant & \textit{2.91\raisebox{-0.5ex}{\scriptsize$\pm$1.35}} & \textbf{2.05\raisebox{-0.5ex}}{\scriptsize$\pm$1.17} & 2.01\raisebox{-0.5ex}{\scriptsize$\pm$1.19} & 2.02\raisebox{-0.5ex}{\scriptsize$\pm$1.18} & 1.94\raisebox{-0.5ex}{\scriptsize$\pm$1.16} & 1.91\raisebox{-0.5ex}{\scriptsize$\pm$1.14} \\
The music sounds natural and fluent & \textit{3.11\raisebox{-0.5ex}{\scriptsize$\pm$1.34}} & 2.11\raisebox{-0.5ex}{\scriptsize$\pm$1.16} & 2.11\raisebox{-0.5ex}{\scriptsize$\pm$1.17} & \textbf{2.16\raisebox{-0.5ex}}{\scriptsize$\pm$1.20} & 2.04\raisebox{-0.5ex}{\scriptsize$\pm$1.13} & 2.02\raisebox{-0.5ex}{\scriptsize$\pm$1.12} \\
The music conveys some emotion & \textit{2.89\raisebox{-0.5ex}{\scriptsize$\pm$1.34}} & \textbf{2.11\raisebox{-0.5ex}}{\scriptsize$\pm$1.21} & 2.03\raisebox{-0.5ex}{\scriptsize$\pm$1.19} & 2.03\raisebox{-0.5ex}{\scriptsize$\pm$1.18} & 1.92\raisebox{-0.5ex}{\scriptsize$\pm$1.12} & 1.94\raisebox{-0.5ex}{\scriptsize$\pm$1.15} \\
The rhythm is consistent & \textit{3.22\raisebox{-0.5ex}{\scriptsize$\pm$1.32}} & \textbf{2.15\raisebox{-0.5ex}}{\scriptsize$\pm$1.20} & 2.11\raisebox{-0.5ex}{\scriptsize$\pm$1.17} & 2.11\raisebox{-0.5ex}{\scriptsize$\pm$1.16} & 2.01\raisebox{-0.5ex}{\scriptsize$\pm$1.13} & 2.00\raisebox{-0.5ex}{\scriptsize$\pm$1.13} \\
The music has a clear structure or repeated segments & \textit{3.61\raisebox{-0.5ex}{\scriptsize$\pm$1.39}} & \textbf{1.95\raisebox{-0.5ex}}{\scriptsize$\pm$1.08} & 1.78\raisebox{-0.5ex}{\scriptsize$\pm$0.95} & 1.76\raisebox{-0.5ex}{\scriptsize$\pm$0.99} & 1.64\raisebox{-0.5ex}{\scriptsize$\pm$0.90} & 1.70\raisebox{-0.5ex}{\scriptsize$\pm$0.86} \\
The music shows a recognizable style & \textit{3.24\raisebox{-0.5ex}{\scriptsize$\pm$1.38}} & \textbf{1.95\raisebox{-0.5ex}}{\scriptsize$\pm$1.12} & 1.79\raisebox{-0.5ex}{\scriptsize$\pm$1.08} & 1.69\raisebox{-0.5ex}{\scriptsize$\pm$1.09} & 1.66\raisebox{-0.5ex}{\scriptsize$\pm$1.00} & 1.73\raisebox{-0.5ex}{\scriptsize$\pm$1.01} \\
The music exhibits tonal coherence & \textit{3.81\raisebox{-0.5ex}{\scriptsize$\pm$1.30}} & \textbf{2.38\raisebox{-0.5ex}}{\scriptsize$\pm$1.20} & 2.16\raisebox{-0.5ex}{\scriptsize$\pm$1.11} & 1.99\raisebox{-0.5ex}{\scriptsize$\pm$1.18} & 2.08\raisebox{-0.5ex}{\scriptsize$\pm$1.12} & 2.10\raisebox{-0.5ex}{\scriptsize$\pm$1.12} \\
The harmonic progression is natural & \textit{3.30\raisebox{-0.5ex}{\scriptsize$\pm$1.31}} & \textbf{2.17\raisebox{-0.5ex}}{\scriptsize$\pm$1.15} & 2.11\raisebox{-0.5ex}{\scriptsize$\pm$1.18} & 1.97\raisebox{-0.5ex}{\scriptsize$\pm$1.24} & 1.92\raisebox{-0.5ex}{\scriptsize$\pm$1.12} & 2.03\raisebox{-0.5ex}{\scriptsize$\pm$1.16} \\
The music exhibits melodic motivation & \textit{3.46\raisebox{-0.5ex}{\scriptsize$\pm$1.37}} & \textbf{2.40\raisebox{-0.5ex}}{\scriptsize$\pm$1.26} & 2.16\raisebox{-0.5ex}{\scriptsize$\pm$1.19} & 1.99\raisebox{-0.5ex}{\scriptsize$\pm$1.22} & 1.99\raisebox{-0.5ex}{\scriptsize$\pm$1.13} & 2.13\raisebox{-0.5ex}{\scriptsize$\pm$1.16} \\
The music sounds novel or original & \textit{2.72\raisebox{-0.5ex}{\scriptsize$\pm$1.33}} & \textbf{2.04\raisebox{-0.5ex}}{\scriptsize$\pm$1.21} & 2.00\raisebox{-0.5ex}{\scriptsize$\pm$1.19} & 1.97\raisebox{-0.5ex}{\scriptsize$\pm$1.16} & 1.90\raisebox{-0.5ex}{\scriptsize$\pm$1.13} & 1.90\raisebox{-0.5ex}{\scriptsize$\pm$1.14} \\
The music left a strong impressive & \textit{2.69\raisebox{-0.5ex}{\scriptsize$\pm$1.36}} & \textbf{1.98\raisebox{-0.5ex}}{\scriptsize$\pm$1.18} & 1.95\raisebox{-0.5ex}{\scriptsize$\pm$1.16} & 1.95\raisebox{-0.5ex}{\scriptsize$\pm$1.14} & 1.83\raisebox{-0.5ex}{\scriptsize$\pm$1.10} & 1.87\raisebox{-0.5ex}{\scriptsize$\pm$1.12} \\
The music reminds me of some experiences & \textit{2.53\raisebox{-0.5ex}{\scriptsize$\pm$1.33}} & \textbf{1.94\raisebox{-0.5ex}}{\scriptsize$\pm$1.16} & 1.91\raisebox{-0.5ex}{\scriptsize$\pm$1.14} & 1.90\raisebox{-0.5ex}{\scriptsize$\pm$1.12} & 1.81\raisebox{-0.5ex}{\scriptsize$\pm$1.09} & 1.85\raisebox{-0.5ex}{\scriptsize$\pm$1.12} \\
I like the music & 
\textit{2.64\raisebox{-0.5ex}{\scriptsize$\pm$1.34}} & \textbf{1.97\raisebox{-0.5ex}}{\scriptsize$\pm$1.17} & 1.94\raisebox{-0.5ex}{\scriptsize$\pm$1.15} & 1.92\raisebox{-0.5ex}{\scriptsize$\pm$1.13} & 1.82\raisebox{-0.5ex}{\scriptsize$\pm$1.09} & 1.88\raisebox{-0.5ex}{\scriptsize$\pm$1.12} \\
\bottomrule
\end{tabular}
\caption{Subjective Evaluation Scores (Mean ± Std), where \textit{Reference} denotes the original samples from the dataset composed by human.}
\label{tab:subjective_scores}
\end{table*}
\subsubsection{Assessment of SNN Models on Core Metrics}
Table~\ref{tab:subjective_scores} shows the total result of subjective evaluation. Specially, figure~\ref{Fig:total-group} (A)–(D) and present the performance of the five SNN models on Q1, Q3, Q11, and Q12 (defined in Table~\ref{tab:subjective_metrics}), which focus musicality, emotional expressiveness, and cognitive perception. Overall, music composed by humans consistently received higher ratings with all mean scores above $2.50$, with score distributions primarily concentrated in the $2$–$4$ range. The total details can be seen in Table A9 in the Appendix. In contrast, SNN-generated music exhibited strong clustering around lower ratings (approximately $1.0$), underscoring the limitations of current SNN models in these dimensions.

Specifically, for music impression and autobiographical association (Q11 and Q12), human-composed pieces achieved the highest average ratings ($2.53$ and $2.69$, respectively), with broader distributions extending toward higher scores ($3$–$4$). Conversely, SNN-generated samples remained tightly clustered around lower scores (mean range: $1.81-1.98$), reflecting limited perceived cognitive engagement.

Among the models, \textbf{S-Transformer} demonstrated the best overall performance across these metrics (Q1 = $2.05$, Q3 = $2.11$, Q11 = $1.94$, Q12 = $1.98$), followed closely by S-GAN, which showed slightly greater variability in listener responses. However, all model-generated results differed significantly from human compositions ($p < 0.001$ across all metrics), as confirmed by Tukey HSD tests (see Tables A7–A9 in the Appendix).

\subsubsection{Variance Across Listener Groups}\label{variance}
Figure~\ref{Fig:total-group} (E)–(F) compares the subjective evaluation scores across three listener groups on four key metrics. Given that S-Transformer consistently achieved the best performance among the five models on these metrics, we use it as a representative example to analyze inter-group variance in subjective perception. For all metrics, human-composed music consistently received higher average ratings across all groups. However, clear inter-group differences emerge. 

An interesting observation is that the amateur group shows the lowest ratings for AI-generated music (Q1 mean = $1.56$, Q3 = $1.62$, Q11 = $1.33$ and Q12 = $1.36$ ), indicating a lower tolerance or stricter judgment, even compared to expert groups. In contrast, for human-composed music, their scores were predominantly distributed in the higher range (Q1 mean = $3.59$, Q3 = $3.68$), indicating a clear preference and stronger positive reception toward human-created pieces. This contrast suggests that amateur participants may exhibit a more polarized attitude, showing both greater appreciation for human compositions and greater skepticism toward AI-generated music.

For the expert group, it's interesting that this group exhibits the relatively higher scores (Q1 mean = $2.31$, Q3 = $2.37$, Q11 = $2.16$ and Q12 = $2.03$) on the same pieces generated by S-Transformer. Furthermore, in terms of musical impression, their ratings for human-composed music were primarily concentrated in the high range ($3$–$4$), whereas the amateur group exhibited a more polarized rating pattern, with scores frequently falling at the extremes (either 1 or 5). A similar trend was observed in responses to personal experience. These results suggest a more open and nuanced evaluative attitude among experts, which may stem from their broader musical exposure and greater familiarity with diverse musical styles.

Listeners in the normal group appeared less sensitive to most musical features, as reflected by their relatively moderate and less variable ratings across all metrics. On S-Transformer-generated music, their scores on Q1 (musicality) and Q3 (emotional expression) averaged $2.24$ and $2.28$, respectively, higher than those from the amateur group but lower than the expert group. Similarly, their evaluations of human-composed music were evenly distributed across the full rating scale (1 to 5). This pattern suggests that the normal group may adopt a more intuitive and less analytical listening strategy.
\subsubsection{Turing Test}
Since the listening process was conducted in a blind manner, the Turing test serves as a classical and crucial component of the overall evaluation. Figure~\ref{Fig:total-group}(I)-(K) presents the Turing test results for the five SNN models across the three participant groups. Overall, participants achieved an average accuracy of 66.83\% in distinguishing between human- and AI-composed music, suggesting a general perceptual ability to differentiate the two. 

Listener groups also exhibited distinct patterns. Participants in the \textbf{normal} group showed limited sensitivity to AI-generated compositions, achieving an overall accuracy rate of only \textbf{58.55\%} in identifying the origin of the music pieces. Furthermore, for both human-composed pieces and those generated by all five SNN models, their accuracy rates remained below 50\%, indicating considerable confusion across sources. These results suggest a reduced ability to differentiate between human- and AI-composed music, potentially due to a lack of specialized musical knowledge or evaluative criteria. 

In contrast, participants in the \textbf{amateur} group reached an accuracy rate \textbf{82.85\%}, the highest among the three groups. Notably, they correctly identified all SNN-generated samples with accuracies exceeding \textbf{59\%}, demonstrating a significantly high sensitivity to AI-generated compositions. 

The \textbf{expert} group achieved an overall accuracy rate of \textbf{78.67\%}. However, an interesting phenomenon is that they exhibited the lowest accuracy in identifying human-composed music, with a rate of only \textbf{28.2\%}. This may indicate that expert participants tend to apply stricter criteria when judging musical quality, due to their extensive musical exposure and high expectations for structural and expressive complexity, potentially leading them to misclassify simpler or less sophisticated human-composed pieces, even forming the reverse Turing bias.

\begin{figure}[t]
\centering
\includegraphics[width=1.0\columnwidth]{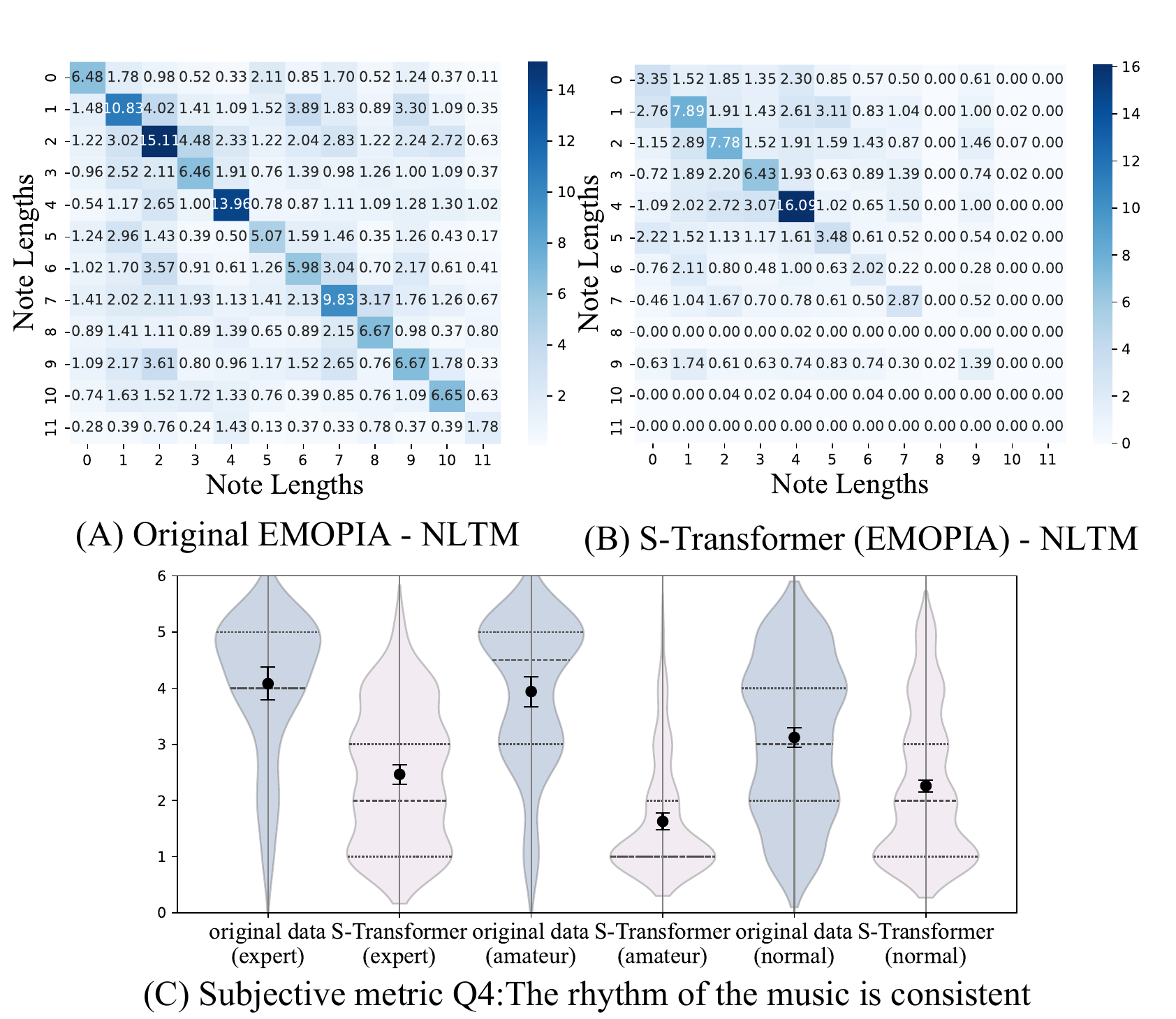} 
\caption{Comparison of objective and subjective metrics for rhythm evaluation on EMOPIA dataset, (A) and (B) present NLPM results on the original data and S-Transformer-generated samples trained on the same dataset, respectively, (C) shows subjective rhythm ratings based on participant perception across three groups.}
\label{Fig:misalignment}
\end{figure}
\subsection{Objective-Subjective Misalignment}
Can objective metrics truly reflect human perceptual judgments? This section analyzes this problem using S-Transformer on EMOPIA dataset.

For pitch-related objective metrics, the \textit{pitch-in-scale rate} evaluates the tonal consistency of a musical piece. The score difference between S-Transformer and human-composed music is minimal ($0.078$, see Figure~\ref{Fig:obj_radar}), indicating that the S-Transformer effectively captures tonal consistency. However, the corresponding subjective evaluation (Q7) shows a statistically significant difference ($p < 0.001$) according to Tukey’s HSD test.

For rhythm-related objective metrics, Figure~\ref{Fig:misalignment}(A) and (B) present the \textit{Note Length Transition Matrices} of the original EMOPIA dataset and S-Transformer outputs, respectively. Additionally, the \textit{Groove consistency} metric in figure~\ref{Fig:rhythm-chord} (A) also shows  strong alignment between generated and original samples. These results suggest that the S-Transformer has successfully captures the rhythm patterns at a statistical level. However, subjective evaluations of rhythm (Q4 in Table~\ref{tab:subjective_metrics}) tell a different story: all three participant groups rated the S-Transformer music significantly lower than the original (mean scores: Expert = $2.47$, Amateur = $1.63$, Normal = $2.26$). This result indicates that the perceived rhythmic quality of the generated music was substantially poorer. 

For harmony, the total difference of \textit{average pitch consonance} between generated and original music is only $0.021$.  Yet, experts rated the harmonic progression of generated samples significantly lower ($-1.131$, $p < 0.001$), indicating that the model failed to produce subjectively convincing harmonic motion.

These discrepancies across pitch, rhythm, and harmony highlight a fundamental misalignment between objective metrics and human perception,  underscoring the necessity of incorporating subjective evaluation in the assessment of music generation models.
\section{Conclusion}
This paper introduces a benchmark and evaluation framework for symbolic music generation using spiking neural networks. By incorporating five representative SNN architectures and five diverse symbolic music datasets, we systematically assess the generative capabilities of these models through both objective metrics and large-scale subjective listening studies. While objective evaluation results reveal that some models, particularly the S-Transformer, could statistically replicate structural features, the subjective results show the gaps between AI- and human-composed music, especially in cognitive aspects proposed in this paper. Furthermore, cross-group analysis demonstrated varied listener sensitivity, highlighting the importance of incorporating diverse human perspectives into evaluation. Our findings emphasize the critical need to move beyond purely statistical metrics in music generation research, underscoring the value of perceptual validation and listener-centered assessment. This work establishes a comprehensive foundation for future research on SNN-based models for symbolic music generation tasks.

\bibliography{aaai2026}
\newpage
\appendix
\onecolumn
\setcounter{table}{0}
\renewcommand{\thetable}{A\arabic{table}}
\captionsetup[table]{labelfont=bf,labelsep=space,name=Table}
\setcounter{figure}{0}
\renewcommand{\thefigure}{A\arabic{figure}}
\captionsetup[figure]{labelfont=bf,labelsep=space,name=Figure}
\section{Appendix}
\begin{table*}[h]
\centering
\begin{tabular}{|c|c|p{2.5cm}<{\centering}|p{2.5cm}<{\centering}|c|p{5.5cm}|}
\hline
\textbf{Dataset} & \textbf{Number} & \textbf{Genre} & \textbf{Annotations} & \textbf{Format} & \textbf{Usage Rationale} \\
\hline
JSB Chorales & 403 & Classical (Chorale) & Chords, Voices & MusicXML & Canonical dataset for polyphonic modeling and harmonic structure learning \\
\hline
POP909 & 909 & Pop & Melody, Chord, Phrase Structure & MIDI & Clean alignment and phrase-level structure suitable for pop-style generation \\
\hline
Lakh MIDI & 176,581 & Mixed (Pop, Jazz, Classical, etc.) & Varied & MIDI & Large-scale corpus with genre diversity and real-world musical distribution \\
\hline
EMOPIA & 1,087 & Piano (Emotional) & Emotion Labels (4-quadrant) & MIDI & Designed for emotion-aware generation with multi-track piano music \\
\hline
XMIDI & 108,023 & Mixed & Emotion, Genre, Instruments & MIDI & Large-scale dataset with expressive metadata and consistent track formatting \\
\hline
\end{tabular}
\caption{Overview of Selected Symbolic Music Datasets Used in Our Benchmark}
\label{Tab:datasets}
\end{table*}

\begin{table*}[h]
\centering
\resizebox{\textwidth}{!}{
\begin{tabular}{l l p{5.3cm} p{5.8cm}} 
\toprule
\textbf{Model} & \textbf{LIF Params} & \textbf{Encoder} & \textbf{Decoder} \\
\midrule

\textbf{S-Transformer} 
& $\tau_m{=}2.0$, $V_{\text{th}}{=}0.6$ 
& Linear(128$\to$256) + LIF + feature embedding. \par
12 Transformer layers, 8 heads, FFN dim=1024. \par
LIF applied after Q/K/V and FFN. \par
LayerNorm applied after LIF.
& Projected Type embedding is concatenated with hidden state. \par
Linear(type+hidden$\to$hidden). \par
Seven feature heads: Linear. \par
Feature heads project hidden representation to features.\\

\midrule
\textbf{S-LSTM} 
& $\tau_m{=}2.0$, $V_{\text{th}}{=}0.6$ 
& Linear(128$\to$1024) + LIF + feature embedding. \par
1-layer LSTM (hidden=256). \par
LIF on outputs.
& Same as S-Transformer. \\

\midrule
\textbf{S-RNN} 
& $\tau_m{=}2.0$, $V_{\text{th}}{=}0.6$ 
& Linear(128$\to$256) + LIF + feature embedding. \par
 Single-layer recurrent unit (hidden=256). LIF on outputs.
& Same as S-Transformer. \\

\midrule
\textbf{S-CNN} 
& $\tau_m{=}2.0$, $V_{\text{th}}{=}0.5$ 
& Linear(128$\to$256) + LIF + feature embedding. \par
Conv1D(128$\to$256) + LIF. \par
& Same as S-Transformer. \\

\midrule
\textbf{S-GAN (Generator)} 
& $\tau_m{=}2.0$, $V_{\text{th}}{=}0.5$ 
& Linear(128$\to$256) + LIF + feature embedding. \par
Conv1D(128$\to$256) + LIF.
& Same as S-Transformer. \\

\textbf{S-GAN (Discriminator)} 
& $\tau_m{=}2.0$, $V_{\text{th}}{=}0.5$ 
& Conv1D(2 layers) + LIF. Final output: Linear(512$\to$1).
& N/A \\

\bottomrule
\end{tabular}%
}
\caption{Architectural Details of Spiking Models}
\label{tab:architecture}
\end{table*}

\begin{table*}[t]
\centering
\begin{tabular}{cp{7cm}p{8cm}}
\toprule
\textbf{Type} & \textbf{Metric} & \textbf{Description} \\
\midrule
\multirow{7}{*}{Pitch} 
& Pitch Count (\textbf{PC}) &  The number of different pitches in a music piece. \\
& Pitch Range (\textbf{PR}) & The span between the highest and lowest pitch in a piece. \\
& Average Pitch Interval(\textbf{PI}) & Average interval between two pitch neighbors. \\
& Pitch Entropy (\textbf{PE}) & The entropy of different pitches in a piece.\\
& Pitch Class Entropy (\textbf{PCE}) & The entropy of 12 pitch class in a music piece.\\
& Pitch-in-scale rate (\textbf{PSR}) & The rate of pitches that belong to the inferred or specified musical scale of a piece.\\
& Polyphony (\textbf{Pp}) & Average number of simultaneous pitches played, indicating texture density (excluding drums). \\
\midrule
\multirow{4}{*}{Rhythm}
& Average inter-onset interval (\textbf{IOI})   & Mean time interval between the onsets of consecutive notes, reflecting rhythmic spacing. \\
& Note Length Transition Matrix (\textbf{NLTM}) & Matrix capturing probabilities of transitions between note lengths, describing rhythmic patterns. \\
& Empty-Beat Rate (\textbf{EBR}) & Calculates the proportion of beats with no active notes (rests or silence). \\
& Groove Consistency (\textbf{GC}) & Evaluates rhythmic regularity by computing cosine similarity across onset vectors in bars. \\

\midrule
\multirow{2}{*}{Harmony}
& Pitch consonance score (\textbf{PCS}) & Average consonance of melody notes to chord tones based on musical intervals within 16th-note windows. \\
& Chord tone to non-chord tone ratio (\textbf{CTnCTR}) & Entropy over chord transition probabilities, reflecting harmonic variety. \\

\bottomrule
\end{tabular}
\caption{Objective evaluation metrics used in our evaluation framework.}
\label{Tab:objective_metrics}
\end{table*}

\begin{table*}[htbp]
\centering
{\footnotesize
\begin{tabular}{ll
    *{7}{c}  
}
\toprule
Dataset & \makecell{Source} 
& \makecell{Average \\ Pitch Interval} 
& \makecell{Pitch \\ Range} 
& \makecell{Pitch \\ Count} 
& \makecell{Polyphony \\ Rate} 
& \makecell{Pitch \\ Entropy} 
& \makecell{Pitch Class \\ Entropy} 
& \makecell{Pitch In \\ Scale Rate} \\
\midrule
\multirow{6}{*}{EMOPIA}
 & S-Transformer & 10.318\textsubscript{±1.190} & 59.652\textsubscript{±7.780} & 40.021\textsubscript{±7.520} & 0.390\textsubscript{±0.086} & 4.876\textsubscript{±0.308} & 3.312\textsubscript{±0.131} & 0.762\textsubscript{±0.057} \\
 & S-LSTM & 11.013\textsubscript{±0.068} & 41.456\textsubscript{±2.826} & 38.565\textsubscript{±2.232} & 0.448\textsubscript{±0.075} & 5.040\textsubscript{±0.108} & 3.529\textsubscript{±0.034} & 0.611\textsubscript{±0.034}  \\
 & S-RNN & 12.113\textsubscript{±0.642} & 44.109\textsubscript{±2.487} & 43.543\textsubscript{±4.174} & 0.468\textsubscript{±0.059} & 5.247\textsubscript{±0.195} & 3.531\textsubscript{±0.102} & 0.650\textsubscript{±0.029}  \\
 & S-GAN & 10.398\textsubscript{±1.241} & 34.260\textsubscript{±2.249} & 24.804\textsubscript{±4.030} & 0.335\textsubscript{±0.123} & 4.418\textsubscript{±0.231} & 3.317\textsubscript{±0.108}  & 0.711\textsubscript{±0.071} \\
 & S-CNN & 12.422\textsubscript{±0.756} & 47.717\textsubscript{±4.266} & 42.043\textsubscript{±3.400} & 0.451\textsubscript{±0.100} & 5.149\textsubscript{±0.112} & 3.515\textsubscript{±0.034}  & 0.640\textsubscript{±0.032} \\
 & Original data & 13.976\textsubscript{±2.321} & 54.391\textsubscript{±7.996} & 31.934\textsubscript{±7.206} & 0.844\textsubscript{±0.142} & 4.495\textsubscript{±0.310} & 2.737\textsubscript{±0.233} & 0.684\textsubscript{±0.265} \\
\midrule
\multirow{6}{*}{JSB}
 & S-Transformer & 11.218\textsubscript{±1.028} & 90.815\textsubscript{±0.642} & 30.157\textsubscript{±2.032} & 0.635\textsubscript{±0.100} & 4.340\textsubscript{±0.078} & 2.978\textsubscript{±0.050}  & 0.869\textsubscript{±0.014} \\
 & S-LSTM & 16.273\textsubscript{±1.669} & 85.947\textsubscript{±0.320} & 30.947\textsubscript{±1.049} & 0.564\textsubscript{±0.103} & 4.683\textsubscript{±0.059} & 3.208\textsubscript{±0.049}  & 0.853\textsubscript{±0.019} \\
 & S-RNN & 10.621\textsubscript{±0.273} & 47.736\textsubscript{±1.772} & 43.131\textsubscript{±1.524} & 0.790\textsubscript{±0.060} & 4.914\textsubscript{±0.027} & 3.328\textsubscript{±0.023}  & 0.825\textsubscript{±0.011} \\
 & S-GAN & 15.043\textsubscript{±3.753} & 72.842\textsubscript{±23.435} & 18.394\textsubscript{±4.386} & 0.308\textsubscript{±0.131} & 4.025\textsubscript{±0.345} & 3.004\textsubscript{±0.212}  & 0.773\textsubscript{±0.083} \\
 & S-CNN & 10.161\textsubscript{±0.583} & 45.394\textsubscript{±2.146} & 38.394\textsubscript{±2.978} & 0.631\textsubscript{±0.106} & 4.874\textsubscript{±0.119} & 3.342\textsubscript{±0.049} & 0.808\textsubscript{±0.025}  \\
 & Original data & 10.922\textsubscript{±0.757} & 32.684\textsubscript{±2.847} & 22.842\textsubscript{±3.090} & 0.998\textsubscript{±0.003} & 4.280\textsubscript{±0.124} & 2.939\textsubscript{±0.120} & 0.863\textsubscript{±0.152}\\
\midrule
\multirow{6}{*}{Lakh MIDI}
 & S-Transformer & 9.524\textsubscript{±0.938} & 61.211\textsubscript{±2.429} & 55.078\textsubscript{±4.670} & 0.422\textsubscript{±0.111} & 5.441\textsubscript{±0.129} & 3.484\textsubscript{±0.033}  & 0.697\textsubscript{±0.042} \\
 & S-LSTM & 9.001\textsubscript{±1.243} & 42.210\textsubscript{±7.837} & 27.605\textsubscript{±7.603} & 0.196\textsubscript{±0.073} & 4.437\textsubscript{±0.369} & 3.329\textsubscript{±0.158}  & 0.756\textsubscript{±0.057} \\
 & S-RNN & 9.912\textsubscript{±0.611} & 43.105\textsubscript{±2.712} & 37.894\textsubscript{±2.062} & 0.415\textsubscript{±0.092} & 4.885\textsubscript{±0.078} & 3.443\textsubscript{±0.366}  & 0.751\textsubscript{±0.023} \\
 & S-GAN & 8.050\textsubscript{±0.903} & 30.921\textsubscript{±2.474} & 23.315\textsubscript{±2.686} & 0.372\textsubscript{±0.112} & 4.263\textsubscript{±0.182} & 3.296\textsubscript{±0.094}  & 0.781\textsubscript{±0.041} \\
 & S-CNN & 8.804\textsubscript{±1.187} & 50.578\textsubscript{±8.791} & 29.710\textsubscript{±6.998} & 0.171\textsubscript{±0.057} & 4.491\textsubscript{±0.254} & 3.344\textsubscript{±0.122}  & 0.754\textsubscript{±0.059} \\
 & Original data & 5.296\textsubscript{±3.932} & 19.236\textsubscript{±14.327} & 12.183\textsubscript{±9.827} & 0.242\textsubscript{±0.335} & 2.790\textsubscript{±1.070} & 2.286\textsubscript{±0.767} & 0.799\textsubscript{±0.256}\\
\midrule
\multirow{6}{*}{POP909}
 & S-Transformer & 7.406\textsubscript{±2.104} & 42.140\textsubscript{±27.815} & 17.180\textsubscript{±3.083} & 0.066\textsubscript{±0.041} & 3.804\textsubscript{±0.244} & 3.123\textsubscript{±0.188}  & 0.702\textsubscript{±0.069} \\
 & S-LSTM & 6.456\textsubscript{±0.720} & 21.940\textsubscript{±1.138} & 20.740\textsubscript{±1.453} & 0.068\textsubscript{±0.029} & 4.165\textsubscript{±0.098} & 3.453\textsubscript{±0.062}  & 0.603\textsubscript{±0.065} \\
 & S-RNN & 6.037\textsubscript{±0.516} & 21.620\textsubscript{±0.797} & 21.700\textsubscript{±0.854} & 0.142\textsubscript{±0.046} & 4.215\textsubscript{±0.080} & 3.500\textsubscript{±0.028}  & 0.629\textsubscript{±0.043} \\
 & S-GAN & 13.193\textsubscript{±2.130} & 68.700\textsubscript{±26.388} & 22.520\textsubscript{±4.332} & 0.329\textsubscript{±0.132} & 4.298\textsubscript{±0.271} & 3.346\textsubscript{±0.152}  & 0.606\textsubscript{±0.087} \\
 & S-CNN & 11.557\textsubscript{±2.985} & 79.520\textsubscript{±11.572} & 21.120\textsubscript{±2.320} & 0.091\textsubscript{±0.041} & 4.162\textsubscript{±0.166} & 3.418\textsubscript{±0.096}  & 0.645\textsubscript{±0.069} \\
 & Original data & 2.450\textsubscript{±0.461} & 19.000\textsubscript{±4.190} & 11.400\textsubscript{±3.085} & 0.732\textsubscript{±0.067} & 4.509\textsubscript{±0.235} & 2.806\textsubscript{±0.163}&0.630 \textsubscript{±0.706} \\
\midrule
\multirow{6}{*}{XMIDI}
 & S-Transformer & 12.485\textsubscript{±2.189} & 42.875\textsubscript{±7.154} & 16.166\textsubscript{±4.709} & 0.125\textsubscript{±0.065} & 3.671\textsubscript{±0.460} & 2.643\textsubscript{±0.346}  & 0.881\textsubscript{±0.069} \\
 & S-LSTM & 11.653\textsubscript{±1.636} & 46.500\textsubscript{±6.570} & 24.200\textsubscript{±7.647} & 0.119\textsubscript{±0.052} & 4.344\textsubscript{±0.445} & 3.183\textsubscript{±0.248}  & 0.745\textsubscript{±0.066} \\
 & S-RNN & 12.772\textsubscript{±0.940} & 47.620\textsubscript{±1.842} & 37.760\textsubscript{±4.052} & 0.234\textsubscript{±0.068} & 4.946\textsubscript{±0.166} & 3.367\textsubscript{±0.068}  & 0.765\textsubscript{±0.039} \\
 & S-GAN & 15.040\textsubscript{±6.406} & 53.600\textsubscript{±25.486} & 11.720\textsubscript{±4.035} & 0.098\textsubscript{±0.100} & 3.377\textsubscript{±0.550} & 2.790\textsubscript{±0.376}  & 0.642\textsubscript{±0.157} \\
 & S-CNN & 13.109\textsubscript{±2.456} & 61.000\textsubscript{±11.447} & 21.700\textsubscript{±7.308} & 0.180\textsubscript{±0.075} & 4.200\textsubscript{±0.483} & 3.081\textsubscript{±0.281}  & 0.782\textsubscript{±0.087} \\
 & Original data & 7.917\textsubscript{±4.641} & 23.980\textsubscript{±19.709} & 16.660\textsubscript{±14.640} & 0.574\textsubscript{±0.267} & 4.381\textsubscript{±0.487} & 2.902\textsubscript{±0.321}& 0.706\textsubscript{0.239±} \\
\bottomrule
\end{tabular}
}
\caption{Pitch-related Metrics (Mean ± Std).}
\label{tab:pitch_metrics}
\end{table*}

\begin{table*}[h]
\centering
{\small
\hspace*{-1.0cm}
\begin{tabular}{ll *{3}{cc}}  
\toprule
\multirow{2}{*}{Dataset} & \multirow{2}{*}{\makecell{Source \\ model-generated / original data}}
& \multicolumn{2}{c}{Average IOI} 
& \multicolumn{2}{c}{Note Length Transition} 
& \multicolumn{2}{c}{Empty Beat Rate} 
\\
\cmidrule(lr){3-4} \cmidrule(lr){5-6} \cmidrule(lr){7-8}
 & & Mean & Std & Mean & Std & Mean & Std \\
\midrule
\multirow{6}{*}{EMOPIA}
 & S-Transformer &  0.307     &   0.082    &    0.974   &   2.241    &   0.165    & 0.071      \\
 & S-LSTM        &   0.231    &     0.384  &    1.467   &   2.551    &    0.183   &     0.068  \\
 & S-RNN         &  0.186     &     0.037  &    3.372   &   4.706    &  0.185     &     0.043  \\
 & S-GAN         &    0.244   &     0.160  &    0.375   &  0.795     &  0.256     &     0.112  \\
 & S-CNN         &     0.215  &     0.058  &    1.151   &  1.841     &    0.188   &     0.075  \\
 & Original data &   0.127    &     0.044  &    1.875   &  5.126     &   0.004    &     0.006  \\
\midrule
\multirow{6}{*}{JSB}
 & S-Transformer &      0.052 & 0.017      &    2.881   &  17.354     &0.139       &     0.508  \\
 & S-LSTM        &  0.091     &     0.028  &    2.025   &   8.790    &  0.112     &     0.053  \\
 & S-RNN         &  0.038     &     0.007  &    8.966   &   31.732    & 0.061      &    0.028   \\
 & S-GAN         &      0.301 &     0.167  &    2.025   &   8.790    &      0.268 &     0.121  \\
 & S-CNN         &  0.081     &     0.023  &    1.428   &   3.855    &  0.160     &     0.054  \\
 & Original data &  0.058     & 0.006       &   1.148    & 10.000      &   0.002    & 0.001      \\
\midrule
\multirow{6}{*}{Lakh MIDI}
 & S-Transformer &      0.097 & 0.034      &    1.967   &  14.687     &   0.239    &    0.090   \\
 & S-LSTM        &  0.705     & 0.364      &    0.646   &   1.855    &  0.414     &     0.106  \\
 & S-RNN         &  0.163     & 0.039      &    1.980   &   9.112    &  0.158     &     0.071  \\
 & S-GAN         &  0.165     &      0.063 &    0.463   &  1.943     &    0.217   &     0.110  \\
 & S-CNN         &  0.641     & 0.186      &    0.729   &  2.020     &  0.384     &     0.077  \\
 & Original data &   1.712    & 8.596      &    1.213   & 19.322      &   0.167    &    0.239   \\
\midrule
\multirow{6}{*}{POP909}
 & S-Transformer &     1.474  & 0.368      &   0.271    &   0.666    &   0.338    & 0.086      \\
 & S-LSTM        &  0.702     &     0.201  &    0.413   &   1.004    &  0.366     &     0.075  \\
 & S-RNN         &  0.507     &     0.150  &    0.814   &  2.201     &  0.352     &     0.072  \\
 & S-GAN         &  0.376     &     0.189  &    0.270   &  0.704     &   0.245    & 0.125      \\
 & S-CNN         &  0.532     &     0.157  &    0.492   &  1.395     &  0.339     & 0.102      \\
 & Original data &  0.157     &     0.025  &    2.243   &   9.657    &   0.012    & 0.015      \\
\midrule
\multirow{6}{*}{XMIDI}
 & S-Transformer &   0.816    &     0.499  &    0.214   &    0.591   &      0.445 & 0.128      \\
 & S-LSTM        &  0.896     & 0.306      &    0.266   &   0.704    &  0.512     &     0.103  \\
 & S-RNN         &  0.401     &     0.117  &    0.841   &  1.817     &  0.335     &     0.095  \\
 & S-GAN         &  0.803     &     0.591  &    0.102   & 0.420      &      0.439 &     0.158  \\
 & S-CNN         &      0.696 & 0.248      &    0.217   &   0.618    &  0.395     &     0.108  \\
 & Original data &  0.091     &     0.087  &    8.909   &  10.737     &   0.247    &    0.136   \\
\bottomrule
\end{tabular}
}
\caption{Rhythm-related Metrics}
\end{table*}

\begin{table*}[htbp]
\centering

\begin{tabular}{ll *{3}{cc}}  
\toprule
\multirow{2}{*}{Dataset} & \multirow{2}{*}{\makecell{Source \\ model-generated / original data}}
& \multicolumn{2}{c}{Average CTnCTR} 
& \multicolumn{2}{c}{Average PCS} 

\\
\cmidrule(lr){3-4} \cmidrule(lr){5-6}
 & & Mean & Std & Mean & Std  \\
\midrule
\multirow{6}{*}{EMOPIA}
 & S-Transformer &  0.519     &  0.054    &   0.794   &   0.071  \\
 & S-LSTM        &   0.360    &     0.040  &    0.748   &   0.039   \\
 & S-RNN         &  0.344     &     0.022  &    0.743   &   0.038   \\
 & S-GAN         &    0.376   &     0.127  &    0.776   &  0.228   \\
 & S-CNN         &     0.346  &     0.040  &    0.743   &  0.055  \\
 & Original data &   0.492    &     0.108  &   0.815  &  0.132     \\
\midrule
\multirow{6}{*}{JSB}
 & S-Transformer &     0.521 & 0.029     &    0.839   &  0.030  \\
 & S-LSTM        &  0.441     &     0.046  &    0.774   &   0.039   \\
 & S-RNN         &  0.433     &     0.018  &   0.786  &   0.020  \\
 & S-GAN         &      0.368 &     0.092  &    0.752   &   0.116  \\
 & S-CNN         &  0.404    &     0.042  &    0.777   &   0.041    \\
 & Original data &  0.536     &   0.093     &   0.837   & 0.094\\
\midrule
\multirow{6}{*}{Lakh MIDI}
 & S-Transformer &     0.467 & 0.032     &   0.759  &  0.042    \\
 & S-LSTM        &  0.420     & 0.057      &    0.749   &   0.093  \\
 & S-RNN         &  0.412     & 0.031    &    0.767   &   0.043  \\
 & S-GAN         &  0.478     &     0.059 &    0.777   &  0.115     \\
 & S-CNN         &  0.416     & 0.063      &    0.756   &  0.070  \\
 & Original data &   0.745   & 0.162     &   0.907  & 0.087   \\
\midrule
\multirow{6}{*}{POP909}
 & S-Transformer &     0.433  & 0.091      &   0.755   &   0.090    \\
 & S-LSTM        &  0.428     &     0.081  &    0.747   &   0.107    \\
 & S-RNN         &  0.440     &     0.064  &    0.765   &  0.091    \\
 & S-GAN         &  0.399    &     0.064  &    0.784   &  0.144      \\
 & S-CNN         &  0.413    &    0.116 &    0.744   &  0.125    \\
 & Original data &  0.818     &     0.188  &   0.834  &   0.155    \\
\midrule
\multirow{6}{*}{XMIDI}
 & S-Transformer &  0.460   &     0.012  &    0.831   &    0.114  \\
 & S-LSTM        &  0.342     & 0.108      &    0.754   &   0.125   \\
 & S-RNN         &  0.382     &     0.052  &    0.768   &  0.064   \\
 & S-GAN         &  0.423    &     0.121  &    0.792   & 0.160   \\
 & S-CNN         &      0.381 & 0.148      &   0.764   &   0.142  \\
 & Original data &  0.721     &    0.188 &    0.834   &  0.155     \\
\bottomrule
\end{tabular}
\caption{Harmony-related Metrics}
\end{table*}

\begin{table*}[htbp]
\centering
{\small
\begin{tabular}{lll}
\toprule
Question & Mean Difference & Significance \\
\midrule
the music is pleasing & -0.8823 & \(p < 0.001\) (Significant) \\
the music sounds natural and smooth & -1.1277 & \(p < 0.001\) (Significant) \\
the music conveys some emotion & -0.7343 & \(p < 0.001\) (Significant) \\
the rhythm of the music is consistent & -1.3916 & \(p < 0.001\) (Significant) \\
the music sounds novel & -0.4334 & \(p < 0.001\) (Significant) \\
the music is impressive & -0.5652 & \(p < 0.001\) (Significant) \\
the music reminds me of some experiences & -0.4984 & \(p < 0.001\) (Significant) \\
I like the music & -0.4255 & \(p < 0.001\) (Significant) \\
the music has a clear structure or repeated phrases & -1.3548 & \(p < 0.001\) (Significant) \\
the music has a distinct style & -0.8514 & \(p < 0.001\) (Significant) \\
the music is tonal & -1.4353 & \(p < 0.001\) (Significant) \\
the music has smooth harmonic progression & -1.1313 & \(p < 0.001\) (Significant) \\
the music contains melodic motifs & -1.0582 & \(p < 0.001\) (Significant) \\
\bottomrule
\end{tabular}
}
\caption{Tukey HSD pairwise comparison between reference samples (from the original dataset) and S-Transformer for the expert group. Negative values indicate lower ratings for S-Transformer.}
\label{tab:tukey_expert}
\end{table*}

\begin{table*}[htbp]
\centering
{\small
\begin{tabular}{lll}
\toprule
Question & Mean Difference & Significance \\
\midrule
the music is pleasing & -1.6203 & \(p < 0.001\) (Significant) \\
the music sounds natural and smooth & -1.8031 & \(p < 0.001\) (Significant) \\
the music conveys some emotion & -1.5333 & \(p < 0.001\) (Significant) \\
the rhythm of the music is consistent & -1.9831 & \(p < 0.001\) (Significant) \\
the music sounds novel & -1.3015 & \(p < 0.001\) (Significant) \\
the music is impressive & -1.4852 & \(p < 0.001\) (Significant) \\
the music reminds me of some experiences & -1.1130 & \(p < 0.001\) (Significant) \\
I like the music & -1.4678 & \(p < 0.001\) (Significant) \\
the music has a clear structure or repeated phrases & -2.0060 & \(p < 0.001\) (Significant) \\
the music has a distinct style & -1.7544 & \(p < 0.001\) (Significant) \\
\bottomrule
\end{tabular}
}
\caption{Tukey HSD pairwise comparison between reference samples (from the original dataset) and S-Transformer for the intermediate group. Negative values indicate lower ratings for S-Transformer.}
\label{tab:tukey_intermediate}
\end{table*}

\begin{table*}[htbp]
\centering
{\small
\begin{tabular}{lll}
\toprule
Question & Mean Difference & Significance \\
\midrule
the music is pleasing & -0.6345 & \(p < 0.001\) (Significant) \\
the music sounds natural and smooth & -0.7463 & \(p < 0.001\) (Significant) \\
the music conveys some emotion & -0.5826 & \(p < 0.001\) (Significant) \\
the rhythm of the music is consistent & -0.7318 & \(p < 0.001\) (Significant) \\
the music sounds novel & -0.5618 & \(p < 0.001\) (Significant) \\
the music is impressive & -0.5326 & \(p < 0.001\) (Significant) \\
the music reminds me of some experiences & -0.4565 & \(p < 0.001\) (Significant) \\
I like the music & -0.5035 & \(p < 0.001\) (Significant) \\
\bottomrule
\end{tabular}
}
\caption{Tukey HSD pairwise comparison between reference samples (from the original dataset) and S-Transformer for the normal group. Negative values indicate lower ratings for S-Transformer.}
\label{tab:tukey_normal}
\end{table*}

\begin{table*}[htbp]
\centering
\hspace*{-0.8cm}
\begin{tabular}{p{5cm}cccccc}
\toprule
Question & Reference & S-Transformer & S-LSTM & S-RNN & S-GAN & S-CNN \\
\midrule
I like the music & 2.64\raisebox{-0.5ex}{\scriptsize$\pm$1.34} & 1.97\raisebox{-0.5ex}{\scriptsize$\pm$1.17} & 1.94\raisebox{-0.5ex}{\scriptsize$\pm$1.15} & 1.92\raisebox{-0.5ex}{\scriptsize$\pm$1.13} & 1.82\raisebox{-0.5ex}{\scriptsize$\pm$1.09} & 1.88\raisebox{-0.5ex}{\scriptsize$\pm$1.12} \\
the music contains melodic motifs & 3.46\raisebox{-0.5ex}{\scriptsize$\pm$1.37} & 2.40\raisebox{-0.5ex}{\scriptsize$\pm$1.26} & 2.16\raisebox{-0.5ex}{\scriptsize$\pm$1.19} & 1.99\raisebox{-0.5ex}{\scriptsize$\pm$1.22} & 1.99\raisebox{-0.5ex}{\scriptsize$\pm$1.13} & 2.13\raisebox{-0.5ex}{\scriptsize$\pm$1.16} \\
the music conveys some emotion & 2.89\raisebox{-0.5ex}{\scriptsize$\pm$1.34} & 2.11\raisebox{-0.5ex}{\scriptsize$\pm$1.21} & 2.03\raisebox{-0.5ex}{\scriptsize$\pm$1.19} & 2.03\raisebox{-0.5ex}{\scriptsize$\pm$1.18} & 1.92\raisebox{-0.5ex}{\scriptsize$\pm$1.12} & 1.94\raisebox{-0.5ex}{\scriptsize$\pm$1.15} \\
the music has a clear structure or repeated phrases & 3.61\raisebox{-0.5ex}{\scriptsize$\pm$1.39} & 1.95\raisebox{-0.5ex}{\scriptsize$\pm$1.08} & 1.78\raisebox{-0.5ex}{\scriptsize$\pm$0.95} & 1.76\raisebox{-0.5ex}{\scriptsize$\pm$0.99} & 1.64\raisebox{-0.5ex}{\scriptsize$\pm$0.90} & 1.70\raisebox{-0.5ex}{\scriptsize$\pm$0.86} \\
the music has a distinct style & 3.24\raisebox{-0.5ex}{\scriptsize$\pm$1.38} & 1.95\raisebox{-0.5ex}{\scriptsize$\pm$1.12} & 1.79\raisebox{-0.5ex}{\scriptsize$\pm$1.08} & 1.69\raisebox{-0.5ex}{\scriptsize$\pm$1.09} & 1.66\raisebox{-0.5ex}{\scriptsize$\pm$1.00} & 1.73\raisebox{-0.5ex}{\scriptsize$\pm$1.01} \\
the music has smooth harmonic progression & 3.30\raisebox{-0.5ex}{\scriptsize$\pm$1.31} & 2.17\raisebox{-0.5ex}{\scriptsize$\pm$1.15} & 2.11\raisebox{-0.5ex}{\scriptsize$\pm$1.18} & 1.97\raisebox{-0.5ex}{\scriptsize$\pm$1.24} & 1.92\raisebox{-0.5ex}{\scriptsize$\pm$1.12} & 2.03\raisebox{-0.5ex}{\scriptsize$\pm$1.16} \\
the music is impressive & 2.69\raisebox{-0.5ex}{\scriptsize$\pm$1.36} & 1.98\raisebox{-0.5ex}{\scriptsize$\pm$1.18} & 1.95\raisebox{-0.5ex}{\scriptsize$\pm$1.16} & 1.95\raisebox{-0.5ex}{\scriptsize$\pm$1.14} & 1.83\raisebox{-0.5ex}{\scriptsize$\pm$1.10} & 1.87\raisebox{-0.5ex}{\scriptsize$\pm$1.12} \\
the music is pleasing & 2.91\raisebox{-0.5ex}{\scriptsize$\pm$1.35} & 2.05\raisebox{-0.5ex}{\scriptsize$\pm$1.17} & 2.01\raisebox{-0.5ex}{\scriptsize$\pm$1.19} & 2.02\raisebox{-0.5ex}{\scriptsize$\pm$1.18} & 1.94\raisebox{-0.5ex}{\scriptsize$\pm$1.16} & 1.91\raisebox{-0.5ex}{\scriptsize$\pm$1.14} \\
the music is tonal & 3.81\raisebox{-0.5ex}{\scriptsize$\pm$1.30} & 2.38\raisebox{-0.5ex}{\scriptsize$\pm$1.20} & 2.16\raisebox{-0.5ex}{\scriptsize$\pm$1.11} & 1.99\raisebox{-0.5ex}{\scriptsize$\pm$1.18} & 2.08\raisebox{-0.5ex}{\scriptsize$\pm$1.12} & 2.10\raisebox{-0.5ex}{\scriptsize$\pm$1.12} \\
the music reminds me of some experiences & 2.53\raisebox{-0.5ex}{\scriptsize$\pm$1.33} & 1.94\raisebox{-0.5ex}{\scriptsize$\pm$1.16} & 1.91\raisebox{-0.5ex}{\scriptsize$\pm$1.14} & 1.90\raisebox{-0.5ex}{\scriptsize$\pm$1.12} & 1.81\raisebox{-0.5ex}{\scriptsize$\pm$1.09} & 1.85\raisebox{-0.5ex}{\scriptsize$\pm$1.12} \\
the music sounds natural and smooth & 3.11\raisebox{-0.5ex}{\scriptsize$\pm$1.34} & 2.11\raisebox{-0.5ex}{\scriptsize$\pm$1.16} & 2.11\raisebox{-0.5ex}{\scriptsize$\pm$1.17} & 2.16\raisebox{-0.5ex}{\scriptsize$\pm$1.20} & 2.04\raisebox{-0.5ex}{\scriptsize$\pm$1.13} & 2.02\raisebox{-0.5ex}{\scriptsize$\pm$1.12} \\
the music sounds novel & 2.72\raisebox{-0.5ex}{\scriptsize$\pm$1.33} & 2.04\raisebox{-0.5ex}{\scriptsize$\pm$1.21} & 2.00\raisebox{-0.5ex}{\scriptsize$\pm$1.19} & 1.97\raisebox{-0.5ex}{\scriptsize$\pm$1.16} & 1.90\raisebox{-0.5ex}{\scriptsize$\pm$1.13} & 1.90\raisebox{-0.5ex}{\scriptsize$\pm$1.14} \\
the rhythm of the music is consistent & 3.22\raisebox{-0.5ex}{\scriptsize$\pm$1.32} & 2.15\raisebox{-0.5ex}{\scriptsize$\pm$1.20} & 2.11\raisebox{-0.5ex}{\scriptsize$\pm$1.17} & 2.11\raisebox{-0.5ex}{\scriptsize$\pm$1.16} & 2.01\raisebox{-0.5ex}{\scriptsize$\pm$1.13} & 2.00\raisebox{-0.5ex}{\scriptsize$\pm$1.13} \\
\bottomrule
\end{tabular}
\caption{Subjective Evaluation Scores (Mean ± Std), where \textit{Reference} denotes human ratings on original samples from the dataset.}
\label{tab:subjective_scores}
\end{table*}

\begin{figure*}[t]
    \centering
    \includegraphics[width=\linewidth]{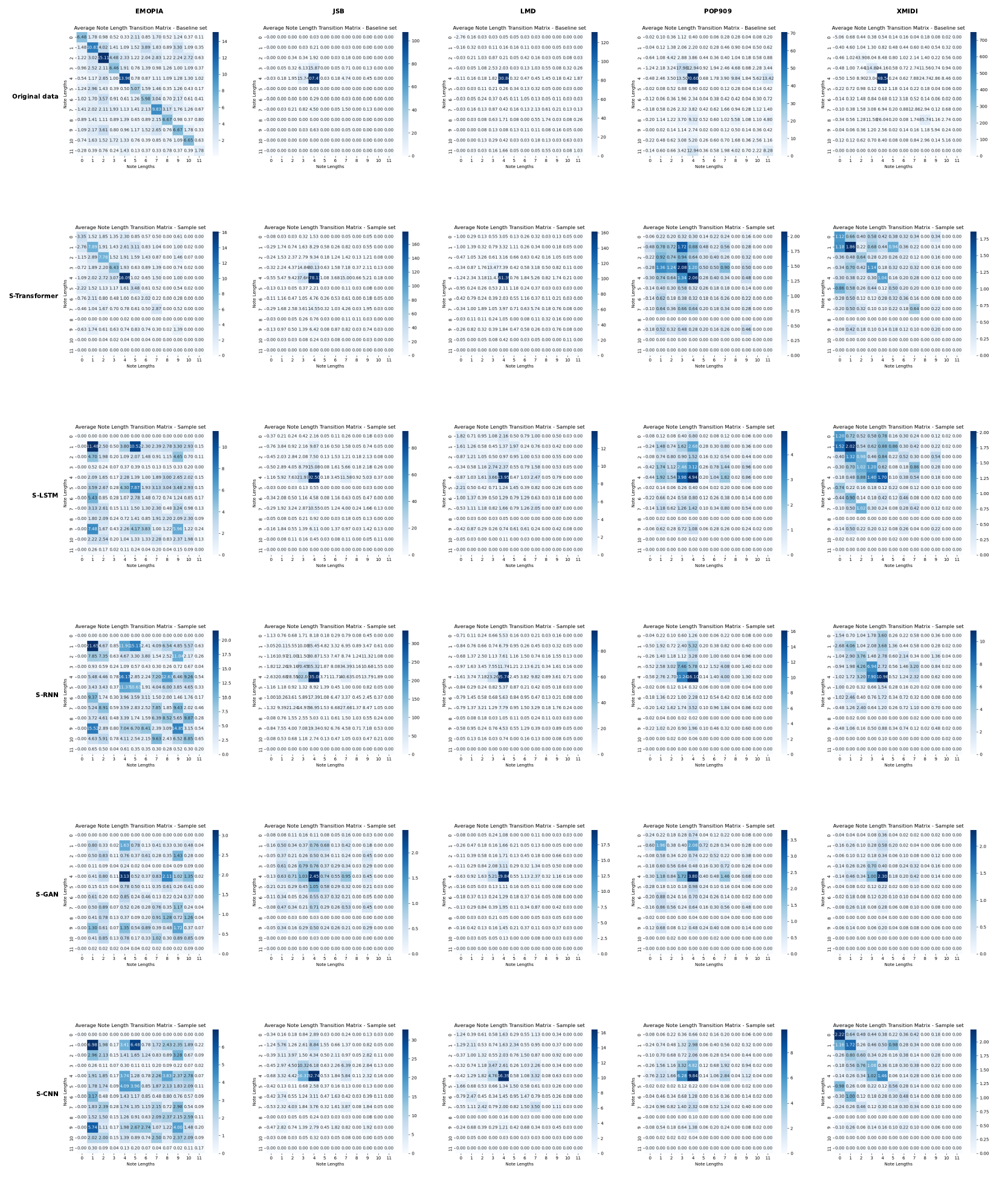}
    \caption{
        Visualization of the \textbf{Note Length Transition Matrix} across different datasets and generated sample sets.
        Each row corresponds to either the original dataset samples or a different generative model,
        while each column corresponds to a dataset.
        Specifically, the first row represents samples from the original datasets.
    }
    \label{fig:merged_heatmaps}
\end{figure*}

\end{document}